\documentclass[12pt,preprint]{aastex}
\usepackage{epsfig}
\newcommand       \Angstrom     {\,{\rm \AA}}

\newcommand       \cm           {\,{\rm cm}}

\newcommand       \erg          {\,{\rm erg}}

\newcommand       \g            {\,{\rm g}}
\newcommand       \K            {\,{\rm K}}

\newcommand       \s            {\,{\rm s}}

\newcommand       \mum          {\,{\rm \mu m}}
\newcommand       \Teff         {T_{\rm eff}}

\newcommand       \simali       {\sim\,}

\newcommand       \Fstar        {F_{\lambda}^{\star}}

\newcommand       \magni        {\,{\rm mag}}
\newcommand       \asic         {\alpha\,{\rm SiC}}
\newcommand       \bsic         {\beta\,{\rm SiC}}
\newcommand       \indexcom     {m}
\newcommand       \indexim      {m^{\prime\prime}}
\newcommand       \indexre      {m^{\prime}}

\newcommand       \omegap       {\omega_{\rm p}}
\newcommand{\figwidth}{4.0in}

\countdef\decade=200
\advance\decade by \year
\countdef\hours=201
\advance\hours by \time
\divide\hours by 60
\countdef\mins=202
\advance\mins by \hours
\multiply\mins by 60
\multiply\hours by 100
\countdef\miltime=203
\advance\miltime by \hours
\advance\miltime by \time
\advance\miltime by -\mins

\shorttitle{On SiC Grains as the Carrier of the 21 Micron Feature}

\begin{document}

\title{
On Silicon Carbide Grains as the Carrier of the 21 Micron Emission
Feature in Post-Asymptotic Giant Branch Stars
     }
\author{B.W. Jiang\altaffilmark{1}, Ke Zhang\altaffilmark{1},
        and Aigen Li\altaffilmark{2}}
\altaffiltext{1} {Department of Astronomy, Beijing Normal University,
                  Beijing 100875, P.R.~China; {\sf bjiang@bnu.edu.cn}}
\altaffiltext{2} {Department of Physics and Astronomy,
                  University of Missouri, Columbia, MO 65211, USA;
                  {\sf LiA@missouri.edu}}

\begin{abstract}
The mysterious 21$\mum$ emission feature seen in
12 proto-planetary nebulae (PPNe) remains unidentified
since its first detection in 1989. Over a dozen of candidate
materials have been proposed within the past decade, but none
of them has received general acceptance. Very recently,
silicon carbide (SiC) grains with impurities were suggested
to be the carrier of this enigmatic feature, based on recent
laboratory data that doped SiC grains exhibit a resonance
at $\simali$21$\mum$. This proposal gains strength from
the fact that SiC is a common dust species in carbon-rich
circumstellar envelopes.
However, SiC dust has a strong vibrational band at
$\simali$11.3$\mum$. We show in this {\it Letter} that in order
to be consistent with the observed flux ratios of the 11.3$\mum$
feature to the 21$\mum$ feature, the band strength of the 21$\mum$
resonance has to be very strong, too strong to be consistent
with current laboratory measurements. But this does not yet
readily rule out the SiC hypothesis since recent experimental
results have demonstrated that the 21$\mum$ resonance of doped
SiC becomes stronger as the C impurity increases.
Further laboratory measurements of SiC dust with
high fractions of C impurity are urgently needed to
test the hypothesis of SiC as the carrier of the 21$\mum$ feature.
\end{abstract}

\section{Introduction\label{sec:intro}}
The so-called ``21$\mum$ feature'' has been identified in
12 proto-planetary nebulae (PPNe; Kwok, Volk, \& Hrivnak 1999)
(and arguably also in in 2 planetary nebulae [PNe] associated
with Wolf-Rayel central stars [Hony, Waters, \& Tielens 2001]
and in 2 highly evolved carbon stars [Volk, Xiong, \& Kwok 2000])
since its first detection in 1989 (Kwok et al. 1989).
This feature has a similar spectral shape and peaks at
the same wavelength ($\simali$20.1$\mum$) in all sources.
The 21$\mum$-feature sources have quite uniform properties:
they are mostly metal-poor, carbon-rich F and G supergiants with
infrared excesses and overabundant s-process elements
(see Kwok et al.\ 1999).

The origin of this feature, however, is still a mystery.
A large number of candidate carriers have been proposed
in the past decade, including hydrogenated fullerenes,
polycyclic aromatic hydrocarbon, hydrogenated amorphous carbon,
diamonds, synthetic carbonaceous macromolecules,
amides (thiourea or urea ${\rm OC\left[NH_2\right]_2}$),
iron oxides (such as Fe$_2$O$_3$ or Fe$_3$O$_4$), SiS$_2$,
oxygen-bearing side groups in coal
(see Kwok et al.\ 1999, Andersen, Posch, \& Mutschke 2005
and references therein), and more recently
titanium carbide (TiC) nanoclusters (von Helden et al.\ 2000),
stochastically-heated silicon core-SiO$_2$ mantle
nanograins (Smith \& Witt 2002; Li \& Draine 2002),
doped SiC (Speck \& Hofmeister 2004),
SiC core-SiO$_2$ mantle grains and iron monoxide FeO
(Posch, Mutschke, \& Andersen 2004).

Among these candidates, TiC nanograins have recently received
much attention, because (1) laboratory spectra of TiC nano-crystals
exhibits a distinct feature at $\sim$~20.1$\mum$, closely resembling
the astronomical 21$\mum$ emission feature both in peak position
and width and in spectral details (von Helden et al.\ 2000),
although bulk TiC does not show any noticeable feature near
20.1$\mum$ (Henning \& Mutschke 2000);
(2) presolar TiC grains are identified in primitive meteorites
as nanometer-sized inclusions embedded in micrometer-sized
presolar graphite grains (Bernatowicz et al.\ 1996).
However, the TiC model has been challenged by Hony et al.\ (2003),
Chigai et al.\ (2003), and Li (2003).\footnote{%
  Hony et al.\ (2003) found that nano TiC must absorb much more
  strongly in the optical and ultraviolet (UV) wavelengths than
  its bulk counterpart in order for the required amount of TiC
  dust not to exceed the (observed) maximum available Ti abundance.
  Chigai et al.\ (2003) found that in order to be consistent with
  the observed flux ratio of the 21$\mum$ and 11.3$\mum$ bands,
  the Ti/Si abundance ratio must be at least 5 times larger than
  the solar abundance ratio. One may argue that the arguments of
  Hony et al.\ (2003) and Chigai et al.\ (2003) do not readily
  rule out the TiC hypothesis because
  (1) as a consequence of the so-called
  {\it electron mean free path limitation} effect,
  the imaginary parts of the dielectric functions of
  small metallic grains
  (and therefore their optical/UV absorptivities)
  are expected to be larger than
  those of their bulk counterparts (see Li 2004);
  (2) there is no reason to compare the solar Ti/Si abundance
  ratio with that of the 21$\mum$-feature sources.
  However, applying the Kramers-Kronig physical principle to
  the TiC hypothesis, Li (2003) readily ruled out the TiC model
  because it was found that this model requires at least 50 times
  more TiC mass than available, no matter how strong
  the UV/optical absorptivity of nano TiC is.
  }

More recently, SiC seems to be an attractive candidate for
the 21$\mum$ feature carrier: Speck \& Hofmeister (2004)
reported the experimental finding that SiC, under certain
circumstances, not only shows the well-known resonance
feature at $\simali$11.3$\mum$, but also a secondary band
which is centered at 20--21$\mum$. This secondary band was
reported to appear only in the $\bsic$-polytype, and nitrogen
or carbon impurities were suspected to favor its occurrence.
This hypothesis gains strength from the fact that
(1) both silicon and carbon are abundant elements; and
(2) SiC is a common dust species in C-rich circumstellar envelopes.
Due to the astrophysical significance of SiC dust,
it deserves a more thorough investigation. In this {\it Letter},
we aim at investigating whether doped-SiC can be a suitable carrier
for the mysterious 21$\mum$ feature by comparing the model-predicted
flux ratio of the 21$\mum$ feature to the 11.3$\mum$ feature
with observed.

\section{Circumstellar SiC Grains and Their Optical Properties\label{sec:opct}}
As early as 1933 Wildt had already suggested that SiC grains
might form in the cool atmospheres of N-type stars.
This suggestion was confirmed 36 years later when Gilman (1969)
and Friedemann (1969) performed thermodynamical equilibrium
calculations and found that SiC grains could condense in
carbon stars. The presence of circumstellar SiC grains
was first revealed by the detection of an emission feature
at 11.5$\mum$ in some carbon stars which was first attributed
to SiC dust by Gilra (1972). The detection of this feature
and its identification as SiC were confirmed by subsequent
observations and interpretations (Treffers \& Cohen 1974;
Merrill \& Stein 1976; Goebel et al.\ 1980;
Little-Marenin 1986; Goebel, Chesseman, \& Gerbaut 1995;
Speck, Barlow, \& Skinner 1997).
The formation of SiC dust in carbon stars has also been
indicated by the identification of presolar SiC grains
in primitive meteorites based on their isotopic anomalies
(Bernatowicz et al.\ 1987).

SiC has $\simali$70 polytypes, which can be divided
into 2 general crystallographic types:
cubic ($\bsic$) and hexagonal ($\asic$).
While presolar SiC grains were predominantly found to
have a cubic lattice structure
(i.e. $\bsic$; Daulton et al.\ 2003),
the 11.3$\mum$ emission feature observed for carbon-rich AGB
stars is best fitted by laboratory spectra of $\asic$
(Baron et al.\ 1987). Speck, Hofmeister, \& Barlow (1999)
suggested that this discrepancy (between meteoritic and astronomical
identifications of the SiC type) is caused by the ``inappropriate
`KBr corrections' made to laboratory spectra of SiC taken
using the KBr matrix method''; they argued that the carrier of
the 11.3$\mum$ feature seen in carbon stars is actually $\bsic$
(instead of $\asic$). More recently, Cl\'ement et al.\ (2003) found
that the laboratory spectra of matrix-isolated $\bsic$ nanoparticles
perfectly match the astronomical 11.3$\mum$ feature.
But Mutschke et al.\ (1999) argued that the 11.3$\mum$ emission
feature itself is not a powerful discriminator of the SiC crystal
type (also see Papoular et al.\ 1998).

Therefore, in this work we will consider both two types
of SiC grains.
We approximate the grains as spherical\footnote{%
  If the grain shape is approximated as a continuous
  distribution of ellipsoids (Bohren \& Huffman 1983),
  both the 11.3$\mum$ feature and the 21$\mum$ feature
  will be significantly broadened. But since at these wavelength
  ranges micron or submicron-sized SiC grains are in the Rayleigh
  regime, the broadening effect is unlikely to differ much
  from one feature to another. Therefore, it is sufficient
  to just consider spherical grains.
  }
and use Mie theory to calculate their absorption
and scattering properties.
The complex index of refraction
$\indexcom(\lambda) = \indexre(\lambda) +
i\,\indexim(\lambda)$ of $\asic$
is taken from Laor \& Draine (1993).\footnote{%
  Laor \& Draine (1983) based their dielectric functions
  ($\epsilon = m^2$)
  in the 11.3$\mum$ wavelength range on those of
  Spitzer, Kleinman, \& Walsh (1959), but broadened
  by a factor of $\simali$16.7. In addition, they
  introduced a ``continuum'' by adding a highly damped
  oscillator.
  }
For $\bsic$, Adachi (1999) compiled the refractive indices
recently determined by various groups in the wavelength range of
$0.13\mum \le \lambda \le 124\mum$, except there was no $\indexim$
data for $0.5\mum \le \lambda \le 0.65\mum$. Largely based on the
data compiled by Adachi (1999), we take the following
``synthetic'' approach: for $\lambda \le 0.13\mum$, we take the
imaginary parts of the refractive indices $\indexim$
from Laor \& Draine (1993);\footnote{%
  At such short wavelengths, the dielectric functions for
  $\asic$ and $\bsic$ should not differ much since they
  just depend on the atomic absorption cross sections of
  the constituent atoms and they are not sensitive to the
  exact crystal structure.
  }
for $0.13\mum \le \lambda \le 0.5\mum$
and $0.65\mum \le \lambda \le 124\mum$,
we take $\indexim$ from Adachi (1999);
for $0.5\mum \le \lambda \le 0.65\mum$,
extrapolation is made from that of Adachi (1999)
at $\lambda \le 0.5\mum$;
for $\lambda > 124\mum$,
we assume
$\indexim(\lambda)=\indexim(124\mu {\rm m})
(124\mu {\rm m}/\lambda)$.
After smoothly joining the adopted $\indexim$, we calculate
the real part $\indexre(\lambda)$
from $\indexim$ through the Kramers-Kronig relation
(Bohren \& Huffman 1983).

If SiC grains are indeed the carrier of the observed 21$\mum$
feature, they must have a resonance at this wavelength
and we should include its contribution to their dielectric
functions. We approximate this contribution in terms of
a single Lorentz oscillator
\begin{equation}\label{eq:epsilon}
\delta\epsilon =
\frac{\omegap^2}{\omega_0^2-\omega^2-i\,\gamma\omega}
\end{equation}
where $\omega\equiv 2\pi c/\lambda$ is the angular frequency
($c$ is the speed of light),
$\omega_0$ is the angular frequency of
the transverse optical mode,
$\gamma =2\pi c\Delta\lambda_0/\lambda_0^2$
is the damping constant ($\lambda_0\approx 20.1\mum$
and $\Delta\lambda_0 \approx 2\mum$
are respectively the peak wavelength
and the FWHM of the 21$\mum$ feature),
and $\omegap$ is the plasma frequency.
For spherical grains in the Rayleigh regime,
the parameters $\omegap$ and $\omega_0$ can be determined
from the feature strength
$\left(Q_{\rm abs}/a\right)$
(Bohren \& Huffman 1983)\footnote{%
  $Q_{\rm abs}$ is the absorption efficiency at $\lambda_0=20.1\mum$
  and $a$ is the grain radius. For grains in the Rayleigh
  approximation (i.e. $2\pi a/\lambda \ll 1$),
  $\left(Q_{\rm abs}/a\right)
  \equiv \left(4\rho/3\right)\kappa_{\rm abs}$
  is independent of $a$, where $\rho\approx 3.22\g\cm^{-3}$
  is the mass density of SiC and $\kappa_{\rm abs}$ is
  the mass absorption coefficient of SiC at 20.1$\mum$.
  In this work, $Q_{\rm abs}/a$ is treated as a free parameter.
  }
\begin{equation}\label{eq:omegap}
\omegap = \frac{1}{3}
\left[\left(\frac{Q_{\rm abs}}{a}\right)
      \frac{3\gamma c}{4}\right]^{1/2}
\left[\epsilon(\infty)+2\right] ~~,
\end{equation}
\begin{equation}\label{eq:omega0}
\omega_0 = \left[\left(\frac{2\pi c}{\lambda_0}\right)^2
- \frac{\omegap^2}{\epsilon(\infty)+2}\right]^{1/2}
\end{equation}
where $\epsilon(\infty)$ is the dielectric function at
$\omega\rightarrow \infty$:
$\epsilon(\infty)\approx 6.6 (6.7)$ for $\asic$ ($\bsic$).

\begin{figure}[ht]
\begin{center}
\epsfig{
        file=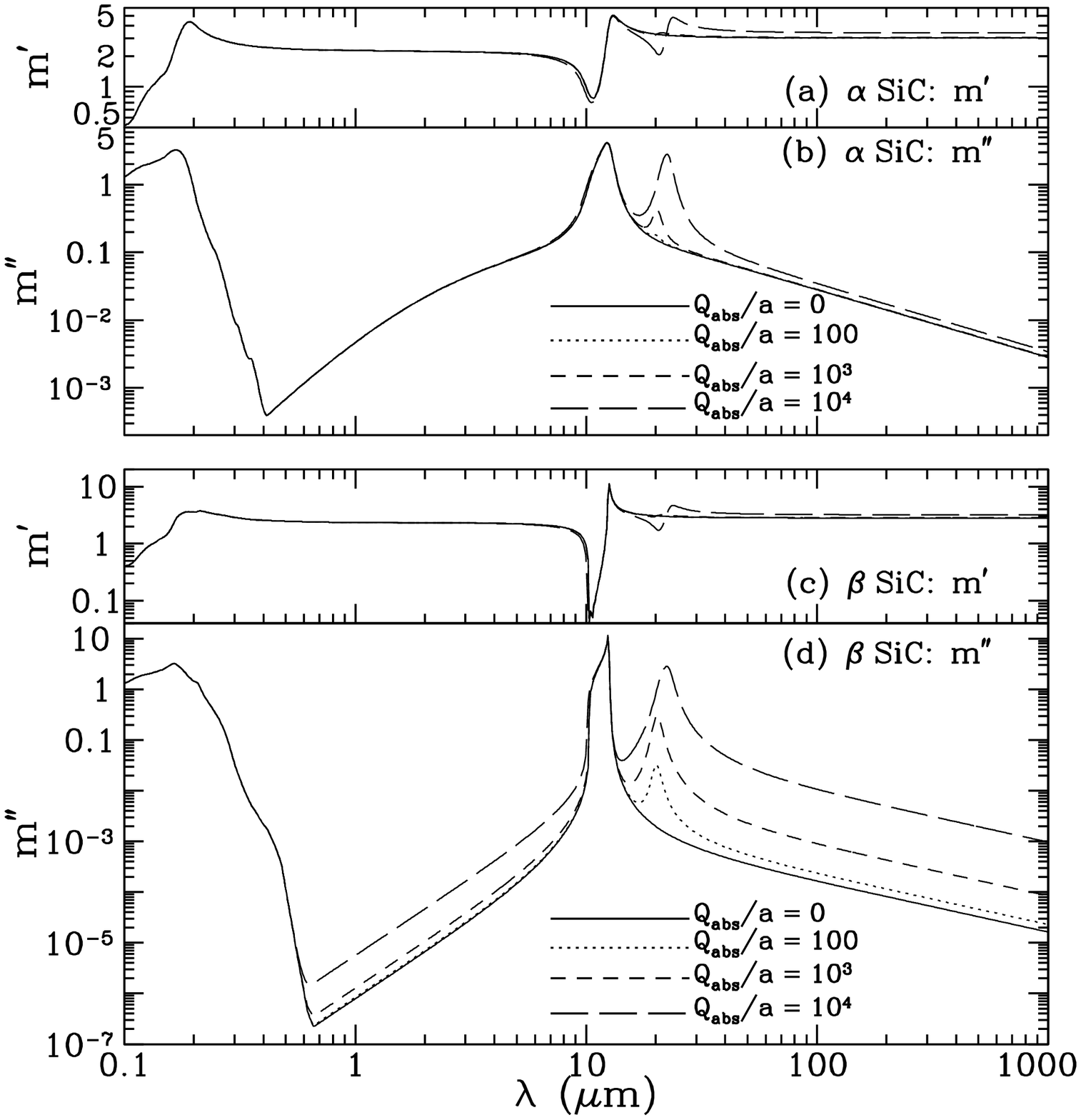, 
        width=\figwidth,angle=0}
\end{center}\vspace*{-1em}
\caption{
        \label{fig:nk}
        Refractive indices
        $m(\lambda)=\indexre(\lambda) + i\,\indexim(\lambda)$
        of $\asic$ (a,b) and $\bsic$ (c,d)
        with various strength for the 21$\mum$ feature:
        $Q_{\rm abs}/a = 0$ (solid lines),
        $Q_{\rm abs}/a = 100\cm^{-1}$ (dotted lines),
        $Q_{\rm abs}/a = 10^3\cm^{-1}$ (short-dashed lines),
        and $Q_{\rm abs}/a = 10^4\cm^{-1}$ (long-dashed lines).
        }
\end{figure}

Unfortunately, Speck \& Hofmeister (2004) did not measure
the absolute strength of the 21$\mum$ feature
$\left(Q_{\rm abs}/a\right)$ for their SiC samples.
We therefore treat $\left(Q_{\rm abs}/a\right)$ as
a free parameter: for a given $\left(Q_{\rm abs}/a\right)$
value, we calculate the contribution of the 21$\mum$ feature
to the dielectric functions of SiC from
eqs.(\ref{eq:epsilon},\ref{eq:omegap},\ref{eq:omega0})
and add this component to the dielectric functions of
Laor \& Draine (1993) for $\asic$ and those described early
in this section for $\bsic$.\footnote{%
  The Lorentz oscillator nature of the 21$\mum$ resonance
  (see eq.[\ref{eq:epsilon}]) guarantees that the new dielectric
  functions also satisfy the Kramers-Kronig relation
  (see Bohren \& Huffman 1983).
  }
For illustrative purpose, in Figure \ref{fig:nk} we plot
the refractive indices of $\asic$ and $\bsic$
with $\left(Q_{\rm abs}/a\right) = 0, 100, 10^3, 10^4\cm^{-1}$.

\section{Results\label{sec:rst}}
The 21$\mum$ sources all have a weak emission feature
at 11.3$\mum$, which is commonly attributed to PAHs
(see Kwok et al.\ 1999). Assuming that all the power
from the 11.3$\mum$ feature is emitted by SiC grains,
one can place constraints on the size and the 21$\mum$
feature strength $\left(Q_{\rm abs}/a\right)$
of SiC dust by comparing the power emitted from
the 11.3$\mum$ feature with that emitted from
the 21$\mum$ feature.
For this purpose, we take HD\,56126 (for which the
dust and gas spatial distributions are well constrained;
see Hony et al.\ 2003, Meixner et al.\ 2004,
Hrivnak \& Bieging 2005) as a testing example.

HD\,56126 (IRAS\,07134+1005), a bright
post-AGB star with a spectral type of F0-5I,
is one of the four 21$\mum$ sources originally discovered
by Kwok et al.\ (1989) and remains the best-studied 21$\mum$
source. The total power emitted from the 11.3$\mum$ feature
and from the 21$\mum$ feature are respectively
$E(11.3\mu {\rm m}) \approx 1.8\times 10^{-11}\erg\s^{-1}\cm^{-2}$
and
$E(21\mu {\rm m}) \approx 1.5\times 10^{-9}\erg\s^{-1}\cm^{-2}$
(Hony et al.\ 2003). Now the question is, with what grain size
and what $\left(Q_{\rm abs}/a\right)$ for the 21$\mum$ feature
one can achieve $E(11.3\mu {\rm m})/E(21\mu {\rm m}) < 0.012$?

Apparently, the observational requirement of
 $E(11.3\mu {\rm m})/E(21\mu {\rm m}) < 0.012$
is best met if the grains are cold (say, with an equilibrium
temperature $\simali$150$\K$) and have a large 21$\mum$ feature
strength $\left(Q_{\rm abs}/a\right)$. Since the further the
grains are away from the central illuminating star, the colder the
grains are, we just need to consider how cold SiC dust can be at
the outer edge of the dusty envelope around HD\,56126. On the
other hand, $\left(Q_{\rm abs}/a\right)$ for the 21$\mum$ feature
can not be arbitrarily large -- it should not be inconsistent
with laboratory measurements
(e.g. see Fig.\,1 in Speck \& Hofmeister 2004).

The equilibrium temperature of a SiC grain
of spherical radius $a$ at the the outer edge
of the dust envelope around HD\,56126 can be
determined by balancing absorption and emission
\begin{equation}\label{eq:T}
\left(\frac{R_\star}{2r_{\rm max}}\right)^2
\int^{\infty}_{0}C_{\rm abs}(a,\lambda)
\Fstar \exp(-A_\lambda/1.086) d\lambda
= \int^{\infty}_{0}C_{\rm abs}(a,\lambda)
4\pi B_\lambda\left(T[a]\right)d\lambda
\end{equation}
where $R_\star \approx 49.2\,r_\odot$ is the stellar radius
($r_\odot$ is the solar radius); $r_{\rm max}\approx 9.3\times
10^{16}\cm$ is the distance from the central star to the outer
edge of the dust envelope (Hony et al.\ 2003);
$C_{\rm abs}(a,\lambda)$ is the absorption cross section
of SiC dust of size $a$ at wavelength $\lambda$;
$T(a)$ is the equilibrium temperature of dust of size $a$
at $r_{\rm max}$; $\Fstar$ is the flux per unit wavelength
(${\rm erg}\s^{-1}\cm^{-2}\mum^{-1}$) at the top of the
illuminating star's atmosphere which is approximated by the Kurucz
(1979) model atmospheric spectrum with $\Teff = 7250\K$ and $\log
g=1.0$; and $A_\lambda$ is the dust extinction which is
represented by the Milky Way standard extinction law with $A_V =
1.0\magni$ (Hony et al.\ 2003).\footnote{%
  This extinction correction is not exact.
  But it does not affect our results since we only
  need to approximately estimate the amount of dust
  attenuation (see Fig.\,1 of Hony et al.\ 2003).
  }
For a SiC grain of a given size $a$ and a given
21$\mum$ feature strength $\left(Q_{\rm abs}/a\right)$,
we calculate its equilibrium temperature from eq.(\ref{eq:T})
and its emission spectrum and then calculate
$E(11.3\mu {\rm m})/E(21\mu {\rm m})$
-- the ratio of the amount of energy emitted from
the 11.3$\mum$ feature to that from the 21$\mum$ feature.

\begin{figure}[ht]
\begin{center}
\epsfig{
        file=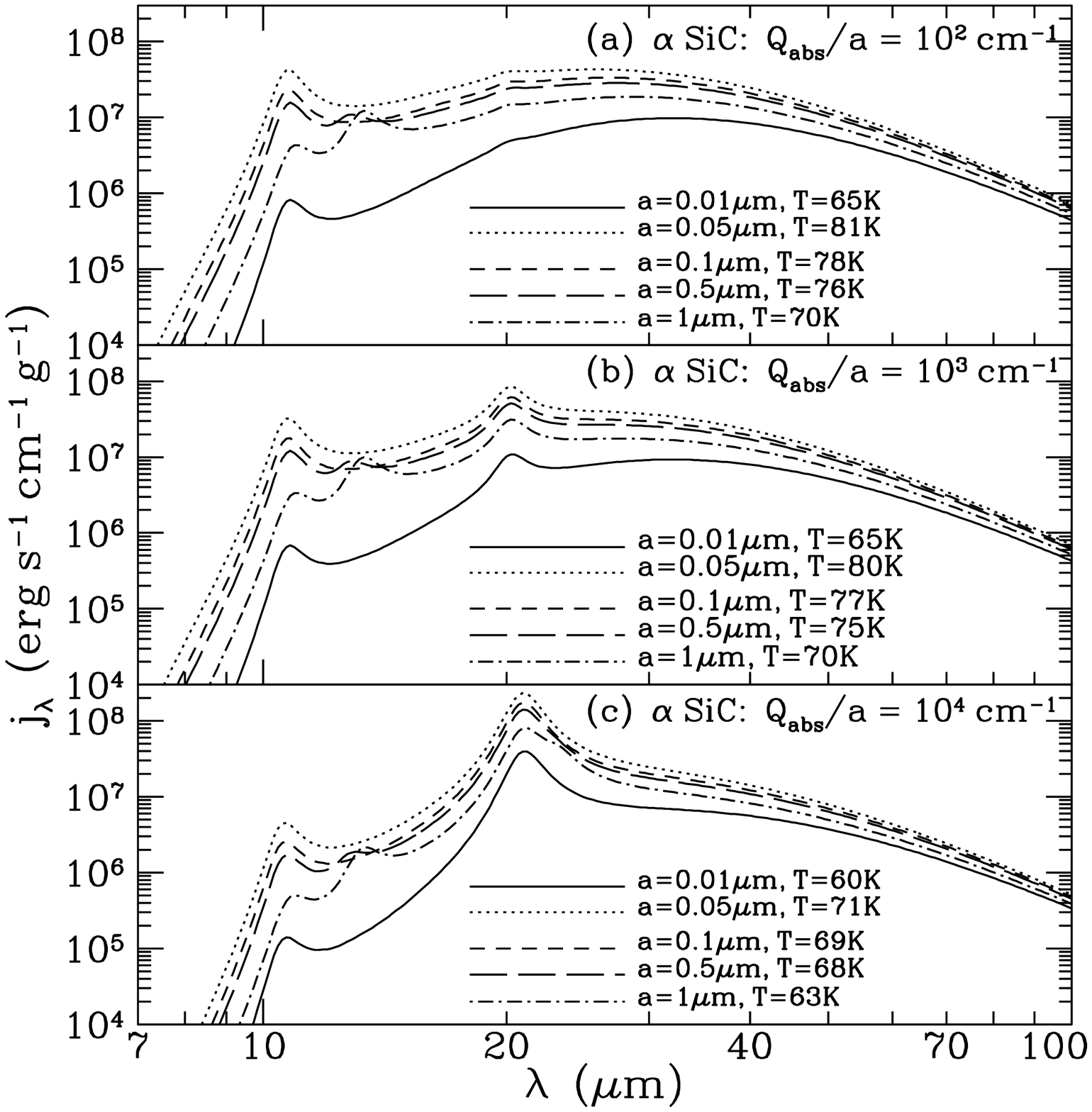, 
        width=\figwidth,angle=0}
\end{center}\vspace*{-1em}
\caption{
        \label{fig:asic}
        Emission spectra for $\asic$ grains
        of sizes $a=0.01\mum$ (solid lines),
        $a=0.05\mum$ (dotted lines),
        $a=0.1\mum$ (short-dashed lines),
        $a=0.5\mum$ (long-dashed lines),
        and $a=1.0\mum$ (dot-dashed lines),
        with $\left(Q_{\rm abs}/a\right) = 100\cm^{-1}$ (a),
        $\left(Q_{\rm abs}/a\right) = 10^3\cm^{-1}$ (b),
        and $\left(Q_{\rm abs}/a\right) = 10^4\cm^{-1}$ (c)
        for the 21$\mum$ feature. The grains are at the outer
        edge of the dust envelope around HD\,56126.
        }
\end{figure}

\begin{figure}[ht]
\begin{center}
\epsfig{
        file=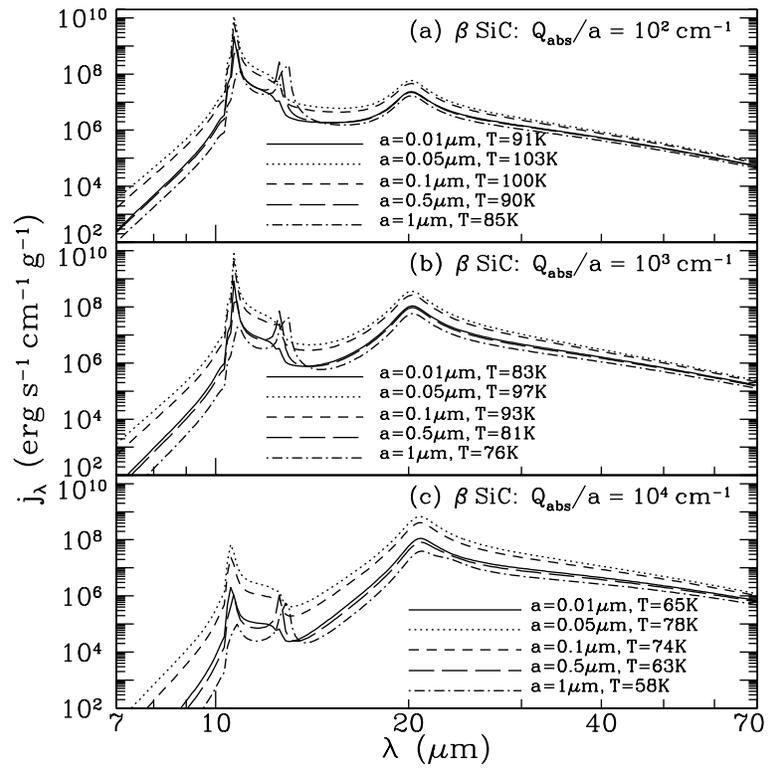, 
        width=\figwidth,angle=0}
\end{center}\vspace*{-1em}
\caption{
        \label{fig:bsic}
        Same as Figure \ref{fig:asic} but for $\bsic$.
        }
\end{figure}

In Figure \ref{fig:asic} we plot the emission spectra of
$\asic$ grains of sizes $a=0.01$, 0.05, 0.1, 0.5 and 1.0$\mum$
with $\left(Q_{\rm abs}/a\right) = 100, 1000, 10^4\cm^{-1}$ for
the 21$\mum$ feature. It is seen that even if one assumes
$\left(Q_{\rm abs}/a\right) = 10^4\cm^{-1}$, the calculated
$E(11.3\mu {\rm m})/E(21\mu {\rm m})$ ratio
($\approx 0.61$ for $a=0.01\mum$
and $\approx 0.41$ for $a=1.0\mum$)
is still much larger than the observed ratio of
$E(11.3\mu {\rm m})/E(21\mu {\rm m})$\,$<$\,0.012.\footnote{%
  Note that $E(11.3\mu {\rm m})/E(21\mu {\rm m})$\,$<$\,0.012
  is already a very generous upper limit since we have attributed
  the entire 11.3$\mum$ emission to SiC, while actually this emission
  must at least partly originate from PAHs.
  }
In order to be consistent with the observation,
one requires $\left(Q_{\rm abs}/a\right) \gg 10^{4}\cm^{-1}$.
This seems unlikely since the required strength for
the 21$\mum$ feature is stronger than that for
the 11.3$\mum$ feature of $\asic$ which is only
$\left(Q_{\rm abs}/a\right) \approx 1.5\times 10^{4}\cm^{-1}$.\footnote{%
   As a matter of fact, in the experimental spectra of
   the $\bsic$ samples of Speck \& Hofmeister (2004),
   the 21$\mum$ resonance is far weaker than
   the 11.3$\mum$ resonance for both bulk and nano materials.
   }
Similar results are obtained for $\bsic$
(see Fig.\,\ref{fig:bsic}).\footnote{%
  In addition to the prominent 11.3$\mum$ band,
  $\bsic$ appears to have another sharp band
  at $\simali$12.6$\mum$. This is caused by the asymmetrical
  nature of its index of refraction (Adachi 1999;
  see Fig.\,\ref{fig:nk}d). If we approximate the 11.3$\mum$
  resonance by a Lorentz oscillator, the secondary peak
  at $\simali$12.6$\mum$ would disappear.
  But this does not affect our conclusion
  since we are considering the total power emitted in
  these features, rather than their peak heights.
  }

\section{Discussion\label{sec:discussion}}
In \S\ref{sec:rst} we only consider grains
large enough to attain an equilibrium temperature.
For small grains with $a$\,$<$\,100$\Angstrom$, they will
undergo single-photon heating: upon absorption of
an energetic photon, they will be heated to a temperature
which is higher than their equilibrium temperature,
and most of the photon energy will be radiated away
at this high temperature. Therefore, the emission
spectrum of a stochastically-heated grain peaks at
shorter wavelengths than that calculated from its
equilibrium temperature (see Draine \& Li 2001).
We thus would expect larger ratios
of $E(11.3\mu {\rm m})/E(21\mu {\rm m})$ for
small grains undergoing single-photon heating
than those considered above, making things even worse.

However, it is still premature to rule out the SiC
hypothesis since the mid-IR spectra of
doped SiC samples of Speck \& Hofmeister (2004)
were obtained at {\it room temperature} for SiC with
a limited fraction of C impurity.
While it is unclear how and to what degree temperatures
will affect the 11.3$\mum$ and 21$\mum$ features of SiC,
it is known that the mid-IR spectra of silicates are
affected by the sample temperature (See Bowey et al.\ 2001).
Moreover, recent experimental results have shown
that the strength of the 21$\mum$ feature increases
as the impurity content increases (Kimura et al.\ 2005a,b).
The required 21$\mum$ feature strength (see \S\ref{sec:rst})
might be attainable by SiC dust with a rather high level
of impurity. Laboratory measurements of the mid-IR spectra
of heavily doped SiC samples are urgently needed.

Finally, we note that there are over 700 C-rich AGB stars
known to show the 11.3$\mum$ SiC feature from the IRAS
({\it Infrared Astronomical Satellite})
LRS ({\it Low Resolution Spectrometer}) survey,
but none of these stars show the 21$\mum$ feature,
suggesting that the amount of impurity incorporated into
SiC dust must be environment dependent
(e.g. the Si/C ratio, Speck \& Hofmeister 2004),
if doped SiC is indeed the carrier of the 21$\mum$ feature.

In summary, we have examined the recent hypothesis
of doped-SiC as the carrier of the mysterious 21$\mum$
emission feature detected in 12 PPNe. It is found that
doped SiC grains have to have a resonance at $\simali$21$\mum$
too strong to be consistent with current laboratory measurements,
in order for the model-predicted flux ratios of the 11.3$\mum$
feature to the 21$\mum$ feature not in conflict with
the observed values. Admittedly, it is still premature
to discard the SiC hypothesis since recent experimental
results have shown that the strength of the 21$\mum$ resonance
of doped SiC appears to increases with the C impurity content,
suggesting that heavily doped SiC may be able to produce
a sufficiently strong 21$\mum$ resonance.
We call on laboratory measurements of the mid-IR spectra
of SiC with high levels of C impurity.

\acknowledgments We thank S. Adachi for sending us his optical
constants of $\beta$-SiC. We thanks A.C. Andersen, B.J. Hrivnak,
H. Mutschke, Th. Posch and A.K. Speck and the anonymous referee
for very helpful comments. This work is in part supported by the
NSFC Project 10473003 of China, the University of Missouri Summer
Research Fellowship, the University of Missouri Research Board,
and the NASA award P20436.

\clearpage

\clearpage

\clearpage

\end{document}